\documentclass[12pt]{article}
\usepackage{xcolor}
\usepackage{tikz}
\usepackage{pgfplots}
\pgfplotsset{compat=newest}
\begin{document}
\begin{titlepage}
\title{ Matter, Space, and Rishons}
\author{
{Piotr \.Zenczykowski }\footnote{E-mail: piotr.zenczykowski@ifj.edu.pl}\\
{\em Professor Emeritus}\\
{\em The Henryk Niewodnicza\'nski Institute of Nuclear Physics}\\
{\em Polish Academy of Sciences}\\
{\em Radzikowskiego 152,
31-342 Krak\'ow, Poland}\\
}
\maketitle
\begin{abstract}
I rephrase my earlier arguments 
that vital information on the emergence
of space is buried in particle physics, and -
in particular - in the Harari-Shupe (HS) rishon model
of leptons and quarks.
First, it is argued that matter and space should be treated 
more symmetrically than they are in the Standard Model.
Then, a generalization of Born's matter-and-space-relating
concept of reciprocity is introduced. 
A simple analogy between the resulting phase-space 
picture and the original HS model is pointed out. 
It is stressed that in the advocated   
view of the rishon model the concept of ``compositeness"
is completely different from its standard understanding,
implying one-dimensionality of rishons, and 
thus the non-existence of ``preons". 
\\
\end{abstract}

\vfill
\end{titlepage}

\section{Introduction}
In  paper \cite{FS} a conjecture was put forward that
essential information on the  emergence of macroscopic
space should be accessible from the study of 
the properties of matter at the hadronic mass 
and distance scales. 
In particular, it was argued that important clues
are contained in the Harari-Shupe (HS) subparticle (rishon)
 model of leptons and quarks \cite{HS,Zen1}.

Given no experimental indication 
of the existence of ``preons",
the argument ``from the rishon model"
is probably discarded by most readers as irrelevant.   
Such an attitude must stem from the macroscopic 
 conception of ``compositeness" which
dominates in our minds 
forcing the majority to think of rishons 
as of ordinary particles.
Actually, in the phase-space-inspired 
version \cite{Zen1,ZenBook} of
the HS model the words ``to be composed of" 
mean something completely different than 
in the original model \cite{HS}. 
Since, despite earlier arguments (see \cite{ZenGlimpse}), 
this difference in meaning remains essentially unnoticed, 
we argue here anew that
the HS model should {\it not} be treated as a preon model.
I think that its most promising interpretation should
be phrased in the language of phase space and its symmetries.

The main aim of this essay is to 
convince the reader that
rishons should be viewed as strictly 
{\it one-dimensional} ``objects" which, 
therefore, cannot be regarded as subparticles.
We point out also that 
the phase-space-motivated interpretation of the 
rishon model offers an unorthodox 
perspective on quark confinement. 
This perspective suggests a close connection of 
the non-existence of free individual quarks
with the idea of the emergence  
of 3D macroscopic space. 
In this essay we present some of 
the arguments of \cite{FS} 
in a rephrased form. 
We focus on the heuristics, hoping  
that in this way the whole conception 
could be conveyed to an interested reader
in a lucid and digestible form.\\
 
\section{Matter in Space}
Our current understanding of the small structures of the world 
is summarized by the Standard Model (SM) of elementary particles. 
Since we want to address the meaning of particle
``compositeness", we start with a brief description of
the relevant underlying concepts.
Among them 
an important ingredient  
is our everyday conception of the world 
as a 3D container through which  spatially extended 3D ``real
material things" move as time flows.
These ``things" 
are characterized by physical concepts, abstracted 
from detailed observations 
of the macroscopic world and best defined
for freely moving objects.
In addition to the notion of a thing's 3D location
specified by vector ${\bf x} = (x_1,x_2,x_3)$,
these are the concepts of the  
constant-in-time quantities:
energy E, momentum vector 
${\bf p}=(p_1,p_2,p_3)$, and mass m. 
The latter concept links energy and momentum,
 with the relevant relativistic dispersion relation being
(in units in which speed of light $c=1$)
 \begin{equation}
\label{eq2} 
 E^2 = {\bf p}^2+m^2.
\end{equation} 
Note that in this formula - which 
describes the basic property of free
relativistic classical things -
the dependence on the space of locations is totally absent. \\
 
\subsection{Matter and the Reductionist Picture} 
The standard reductionist picture consists of a set
of steps in which 
big ``real" material 3D things are viewed as 
composed of spatially smaller and less massive but
otherwise similar ``real" material 3D things. 
The components are best identified when  
their mutual interactions vanish.
     In our macroscopic world this happens
 when the components are 
spatially well separated from each other.
In order to avoid infinite regress, the reductionist 
procedure has to stop somewhere.
The relevant ``atomicity" assumption postulates then
the existence of some underlying indivisible units of matter. 
In the current theoretical picture of the SM, 
the role of these ``atoms" is played by elementary particles. 
Those particles that can be spatially separated  
are identified by the condition
that the macroscopically motivated classical 
constraint of Eq. (\ref{eq2}) be satisfied.  
 
 As we go down the reductionist ladder,
 the macroscopic classical ``reality" 
of matter slowly dissappears.
For example, in the quantum microscopic world 
the dispersion relation (\ref{eq2}) 
ceases to be exactly satisfied and
the concept of virtual particles appears.
Despite this clear indication of a loss of
the ``reality" of matter,
influenced by our everyday macroscopic experience 
and the time-proven reductionist ideas,
we are continually tempted to ask: 
from what kind of subparticles 
is matter composed at the nuclear level?
at the hadronic level?
at the level of quarks and leptons? 
Yet, as 
the appearance of particle virtuality
 at the classical-to-quantum transition suggests,
 further substantial changes 
in our conception of matter should be expected
at each subsequent major step down the reductionist ladder.
 
 In fact,
many physicists suggested that 
at the level of elementary particles
 the concept of (``real") matter  should be  replaced 
by the abstract concept of symmetry.
When in 1964 Murray Gell-Mann first introduced quarks 
as components of hadrons \cite{Gell-Mann},
he thought of them as of quantum fields satisfying 
an abstract symmetry. 
He was apparently concerned with the danger of treating
quarks as ordinary ``material" particles 
as the last sentence of his original paper demonstrates: 
{\it ``A search for stable quarks of charge -1/3 or +2/3 (...) 
 at the highest energy accelerators would help to 
reassure us of the non-existence of real quarks."} 
Today, the statement that ``hadrons are composed of quarks"
accepts the ``reality" of quarks, though subject to
 a confinement-related correction of the meaning
of hadron ``compositeness". 
Indeed, in the SM, strong interactions between component 
quarks vanish not at large but at small separations. 
May we then consider quarks as truly ``real"? 
Are the correponding quark dispersion
relations not affected by confinement?
 
 The symmetry of particle interactions and the pattern of
lepton-quark generation are probably the most
important ingredients of the SM. Over the years, 
the relevant SM symmetry group  has been identified as
$U(1) \otimes SU(2)_L \otimes SU(3)_C$ with group factors
describing the symmetries of
the electromagnetic, weak, and strong interactions, 
respectively. 
Yet, despite its many particular successes,
the SM leaves unanswered the fundamental question: 
why is the world governed by the 
$U(1) \otimes SU(2)_L \otimes SU(3)_C$ symmetry 
group?\footnote{Simply pointing out some larger symmetry 
group is not sufficient. The relevant symmetry 
should follow from a deep physical principle.} 
What physical principle could explain the relevance of 
this particular symmetry?
Should we seek an explanation of 
this symmetry in terms of ``real" subparticles 
of the reductionist paradigm,
or assume it as a fundamental input 
that does not need any further explanation?
Or perhaps one should look for another kind of explanation?
And what about the pattern of lepton-quark generation?\\

 \subsection{The Rishon Model}
Driven by the repeated successes of 
the reductionist approach, 
the concept of ``explanation" was usually 
understood in particle or subparticle terms.
Prompted by such thinking, a subparticle  
explanation of the basic features of the
SM symmetry group and the lepton-quark generation structure 
was proposed some 40 years ago by Harari and Shupe
in their rishon model of composite leptons and quarks \cite{HS}.

 The HS model assumes 
that each member of a single generation of leptons and quarks
is built from two spin-1/2 subparticles (or their antiparticles): 
the ``rishons" $T$ and $V$ of electric 
charges $Q_T=+1/3$ and $Q_V=0$. 
Due to some unknown confinement mechanism, 
these rishons are combined in {\it ordered} sets of 
three as shown in Table \ref{table1} 
for the upper ($I_3=+1/2$) components 
of weak isospin doublets (eg. $\nu_e,e^-$). 
As charge $Q$ is related to weak isospin $I_3$ 
and hypercharge $Y$ through 
\begin{equation} 
Q = I_3 + Y/2,
\end{equation}
we have $Y_T= +1/3$, $Y_V= -1/3$, and
the rishon structure of lepton and quark charges 
translates into a
corresponding structure of hypercharges. 
 Likewise, the eight states of $I_3 = -1/2$, 
ie. $e^-$,   $d_R$, $d_G$, $d_B$ 
and $\bar{\nu}_e$, $\bar{u}_R$, $\bar{u}_G$,  $\bar{u}_B$  
are composed of rishon antiparticles $\bar{T}$, $\bar{V}$.

 \begin{table}
\caption{Rishon structure of leptons and quarks with a third component of weak
isospin $I_3 = +1/2$}
\begin{center}
\begin{tabular}{l | c c c c | c c c c}
\hline
      & $e^+$ & $\bar{d}_R$ & $\bar{d}_G$ & $\bar{d}_B$   
&     $\nu_e$ &  $u_R$ &  $u_G$ &  $u_B$ \rule{0mm}{6mm}\\
\hline
      & $TTT$ &  $TVV$ &  $VTV$ &  $VVT$ & $VVV$ & $VTT$ & $TVT$ & $TTV$ \rule{0mm}{6mm}\\
Q   &  +1  &  +1/3 & +1/3 & +1/3      &     0  & +2/3 & +2/3 & +2/3 \rule{0mm}{6mm}\\
Y   &  +1  &  -1/3 & -1/3 & -1/3      &     -1 & +1/3 & +1/3 & +1/3 \rule{0mm}{6mm}\\
\hline
\end{tabular}
\end{center}
\label{table1}
\end{table}
  Although the HS model nicely explains both
the appearance of the $U(1)\otimes SU(3)_C$ part
of the SM symmetry group and 
the pattern of lepton-quark generation, 
it exhibits many   
shortcomings that are induced by the assumed 
particle nature of the components
(see \cite{Zen1,ZenBook,ZenGlimpse}).
Regarding these shortcomings as deadly for 
the reductionist 
view of the model, 
one may seek a different explanation 
of the SM symmetries (the $U(1) \otimes SU(3)_C$ symmetry 
group in particular): not in terms of a 
``rishonian" rung
on the reductionist ladder but in terms of a link 
between the quark-lepton rung and some macroscopic 
classical concepts that were not considered
earlier. 
We may recall here Bohr's words:
 {\it ``It should be made clear that this 
theory\footnote{Bohr's model of atom} 
is not intended to explain phenomena 
in the sense in which word `explains' 
has been used in earlier physics. 
It is intended to combine various phenomena, 
which seem not to be connected, 
and  to show that they are connected."}
Why should we not think therefore of 
linking some previously unconnected 
micro and macro concepts? Why should we not try
to link the successful features of the HS model with some 
properties of macroscopic reality?

 \section{Matter and Space}  

In the SM the elementary particles (the ``atoms")
 and the background they move in (the ``container") are treated as 
essentially disjoint concepts.
 This is not so in the thinking that contributed to Einstein's 
creation of General Relativity (GR), in which 
gravitational forces are reduced to aspects of space geometry.
  According to this way of thinking, 
the existence and/or properties of space are induced  
 by (or related to) the existence of matter. For example,
space may be viewed as a structure defined by relations between
chunks of matter.
Consequently,
the SM picture of the ``container" space 
through which material things move 
must be seen as an over-simplified 
description of the situation:
it misses the matter-related nature of space. 
     Thus, the SM appears as a hybrid ``cq" theory 
that mixes in an asymmetric way the
classical macroscopic (space, ``c") 
and the quantum microscopic (matter, ``q") 
aspects of reality \cite{Finkelcq}.
Keeping in mind the basic role played in the SM 
by matter and interaction symmetries, 
we think that, in a future deeper theory, matter and space 
should also be treated symmetrically. 

 \subsection{The Symmetry of Reciprocity}
 
A possible candidate for the matter-space 
symmetry was proposed by Max Born.
He observed \cite{Born15}
that while in formula (\ref{eq2}) the
mass of free physical bodies appears
in association with momentum only
(ie. {\it not} with
position), various
other important physical formulas, 
such as eg. Hamilton's equations of motion,
 the classical expression for the angular momentum 
${\bf J} = {\bf x} \times {\bf p}$ 
as well as the position-momentum quantum commutation 
rules $[x_j, p_j ] =i h$ (or the related classical
position-momentum Poisson brackets)
exhibit exact symmetry under 
the position-momentum interchange.
 
 This symmetry of ``reciprocity" suggests the existence 
of a new physical constant $\kappa$ 
of dimension [momentum/position] 
that permits the expression of momenta 
as {\it proportional} to positions, according to
\begin{eqnarray}
\label{reciprocity}
{\bf p} & = & \kappa {\bf x},\nonumber \\
{\bf x} & = & - \kappa^{-1} {\bf p}.
\end{eqnarray}
Constant $\kappa$ 
does not have much to do with the quantum 
Planck constant $h$ of dimension  
[momentum $\times $ position] 
 which permits the expression of momenta 
as {\it inversely} proportional to positions.
Born speculated that 
 the ordinary concept of mass (and thus that of matter) 
should be generalized and should include
 the classical position variable 
${\bf x}$ (and thus the concept of space), probably 
via some analogue of dispersion formula (\ref{eq2}).
 
Now, reciprocity treats matter and space as symmetry-related
concepts that may be transformed into each other. 
It might be therefore of interest in various 
gravity-related contexts.
Indeed, the role of reciprocity as 
a guiding principle in search
of a proper approach to quantum gravity 
has been pointed out recently 
by Buoninfante \cite{Buoninfante} 
who noticed that reciprocity permits
 to {\it ``incorporate already in flat space  (...) 
a fundamental acceleration scale 
corresponding to a maximal limiting value $a_P$"}.
In fact, limits on acceleration may 
be argued to be more fundamental 
than corresponding limits on distance:
after all, it is acceleration that 
describes the strength of matter-induced gravitational field 
and its connection with the properties of space.

With the Planck (maximal) acceleration 
being 
\begin{equation}
\label{aP}
a_P =c^2/l_P = m_Pc^3/h \approx 2.2 \times 10^{53} ~cm/s^2
\end{equation}
($m_P=5.5 \times 10^{-5}~g$ and $l_P= 4 \times 10^{-33}~cm$ are Planck's mass and length), 
we obtain  \cite{FS}
\begin{equation}
\label{kappaP}
\kappa_P = a_P^2 h/ c^4  
 =4.04 \times 10^{38} ~g/s \equiv \kappa_C= c^3/G.
\end{equation} 
This equation gives 
the maximum (`classical') value of $\kappa$, reached 
at the surface of a Planck-size black hole, 
and defined by two classical constants: 
$c$ and Newton's gravitation constant $G$.

With Einstein's gravitational field equation
involving cosmological constant $\Lambda$ 
in addition to $c$ and $G$,
one may define another (`quantum') 
 value of $\kappa$ \cite{FS}: 
\begin{equation}
 \kappa_Q = h \Lambda = 0.79 \times 10^{-82} g/s,
\end{equation}
which is 120 orders of magnitude smaller than $\kappa_C$, 
and thus may be considered minimal.  
The corresponding value of acceleration 
$a_Q = \sqrt{\kappa_Q c^4/h}=c^2\sqrt{\Lambda} 
\approx 9.8 \times 10^{-8} ~ cm/s^2$
is of the order of MOND acceleration 
$a_M =1.2 \times 10^{-8}~ cm/s^2$ \cite{MOND}.  
Thus, $\kappa_Q$ defines the minimal
value of acceleration $a_Q$ below which departures from
the standard picture of gravitational forces
are expected to appear.\footnote{Such departures 
are in fact observed
in astrophysical settings for $a < a_M$. They are
 well explained in the framework of
modified Newtonian dynamics (MOND) \cite{MOND}.}

When the atomicity postulate is added, 
reciprocity suggests that 
apart from the ``atoms of matter" 
(exhibiting a discrete spectrum of masses 
associated through (\ref{eq2}) 
with momentum vector ${\bf p}$),
 there should be
``atoms of space" (associated with position vector ${\bf x}$). 
Actually, the ``atoms" 
of space need not be as tiny as the ``natural"
unit of Planck length suggests. 
The only essential argument that points
towards this diminutive distance scale 
is provided by the dimensional analysis. 
Yet, such an analysis is not trustworthy as its
results depend on the choice of 
constants considered to be fundamental \cite{Meschini}.
Since Einstein's field equation
involves three universal classical constants: 
$c$, $G$, and $\Lambda$,
 one has to decide
 which two of them
should be selected to supplement $h$
in the dimensional analysis 
of the ``natural" scales for quantum gravity.
The standard Planck's mass and length 
scales $m_P$ 
and $l_P$ 
follow if 
 $h$, $G$, and  $c$ are used.
The choice of $h$,  $G$ 
and  $\Lambda$ instead
leads to the hadronic mass 
 and length scales \cite{FS}:
\begin{eqnarray}
\label{mH}
m_H & = & \left((h^2/G)\sqrt{\Lambda/3}\right)^{1/3} 
\approx  0.35 \times 10^{-24}~g , \\
l_H & \approx & 0.64 \times 10^{-12}~cm, 
\end{eqnarray}
and to a rough estimate of 
the slope $\alpha '$ of hadronic
Regge trajectories \cite{FS}
\begin{equation}
\label{alphaprim}
\alpha ' 
=\left(\frac{3G^2}{h\Lambda}\right)^{1/3}
\approx 55 \times 10^{21}~cm^2/(gs)\approx  N^{1/3} c^2 /\kappa_P.
\end{equation}
Here $N=3 c^3/(h G \Lambda)=1.54 \times 10^{121}$.

The obtained scales are in an order-of-magnitude agreement 
with their hadronic values. 
While this agreement is very good for hadronic mass scale,
the estimated string slope scale of Eq. (\ref{alphaprim}) 
differs from the experimental hadronic
value of $\alpha ' = 0.38\times 10^{21}cm^2/(gs)$
by a factor of 100 or so. Such a factor is quite acceptable
when the large number multiplicative factor of approximately $10^{40}$
between
$\kappa_H \approx (\kappa_C^2\kappa_Q)^{1/3}
=c^2(\frac{h\Lambda}{3G^2})^{1/3} \approx 2.4~g/s$ 
and
$\kappa_P\equiv \kappa_C=4.04 \times 10^{38} ~g/s$
is taken into account \cite{FS}.
It is hard to believe that this double coincidence 
of the estimated and the measured 
hadronic mass and string slope scales 
does not tell us something important
about the hadronic involvement in the 
emergence of ordinary space from a quantum quark layer.

The two sets of alternative scales ($l_P,m_P$ and $l_H,m_H$) estimated
by the dimensional analysis
probably correspond to different
limiting aspects of space-related quantum effects 
(see Fig. 1 in \cite{Zen2022}).

\subsection{Beyond Reciprocity}
With the help of $\kappa$, the two relevant
3D invariants ${\bf p^2}$  
and ${\bf x^2}$ may be combined
to form the matter-and-space symmetric 
expression ${\bf p^2}+{\bf x^2}$ 
(in units in which $\kappa =1$).
In this way, the concepts of the 3D matter 
(momentum) and 3D position spaces
get unified into that of a 6D phase-space in which 
 matter and space  variables may be treated on more
equal footing \footnote{Thus, the principle of reciprocity
provides the reason for the extension of rotational symmetry from 3D to 6D.}. 

If $\kappa$ exists, then
for any object in our 3D world
each of the three perpendicular directions
is associated with a pair
of physically different (position and momentum)
variables of the same dimension, 
which, consequently, may be freely exchanged.
Thus, we may generalize the symmetry of reciprocity 
to that of an exchange (permutation) 
symmetry for individual momentum 
and position coordinates, 
ie.
\begin{eqnarray}
\label{exchange}
p_i & = & \kappa x_i,\nonumber \\
x_i & = & - \kappa^{-1} p_i,
\end{eqnarray}
with specific fixed $i$,
as shown in Table \ref{table2}. 

\begin{table}[t] 
\caption{Permutations of individual 
momentum and position components 
of 6D phase space vector
$({\bf p}; {\bf x}) = (p_1, p_2, p_3; x_1, x_2, x_3)$}
\begin{center}
\begin{tabular}{l c r}
Sector & Exchange & Momentum; position\rule{0mm}{6mm}\\
\hline
Odd number of exchanges & & \rule{0mm}{6mm}\\
\hline
Blue		& $x_3 \leftrightarrow p_3$ &    ($p_1, p_2, x_3; x_1, x_2, p_3$) \rule{0mm}{6mm}\\
Green		& $x_2 \leftrightarrow p_2$ &    ($p_1, x_2, p_3; x_1, p_2, x_3$) \rule{0mm}{6mm}\\
Red		& $x_1 \leftrightarrow p_1$ &    ($x_1, p_2, p_3; p_1, x_2, x_3$) \rule{0mm}{6mm}\\
Reciprocity	& ${\bf x}\leftrightarrow {\bf p}$ &($x_1, x_2, x_3; p_1, p_2, p_3$) \rule{0mm}{6mm}\\
\hline
Even number of exchanges &&\rule{0mm}{6mm}\\
\hline
Red		&$(x_3, x_2) \leftrightarrow (p_3, p_2)$  & 
$({\bf p}_R; {\bf x}_R) = (p_1, x_2, x_3; x_1, p_2, p_3)$ \rule{0mm}{6mm}\\
Green		&$(x_1, x_3) \leftrightarrow (p_1, p_3)$  &
$({\bf p}_G; {\bf x}_G) = (x_1, p_2, x_3; p_1, x_2, p_3)$ \rule{0mm}{6mm}\\
Blue		&$(x_2, x_1) \leftrightarrow (p_2, p_1)$  &
$({\bf p}_B; {\bf x}_B) = (x_1, x_2, p_3; p_1, p_2, x_3)$ \rule{0mm}{6mm}\\
Identity	&${\bf x} \to {\bf x}$, ${\bf p} \to {\bf p}$       &
($p_1, p_2, p_3; x_1, x_2, x_3$) \rule{0mm}{6mm}\\
\hline
\end{tabular}
\end{center}
\label{table2}
\end{table}

The four even and four odd sets of exchanges suggest
a generalization of the classical 
association (\ref{eq2}) 
of the standard concept of mass (or particle) with momentum
(ie. the dispersion relation involving 
vector $(p_1, p_2, p_3)$) 
to eight such associations (involving  
the eight ordered triplets 
on the left in the last column of Table \ref{table2}, ie.
$(p_1, p_2, x_3)$, $(p_1, x_2, p_3)$, ..., $(x_1, x_2, x_3)$; 
$(p_1, x_2, x_3)$, ..., $(p_1, p_2, p_3)$,
each playing the role of 
the ordinary momentum vector $(p_1, p_2, p_3)$).
 
Note the similarity between 
the eight sets of generalized momentum
triplets shown in Table \ref{table2}
and the eight sets of ordered three-rishon 
states shown in Table \ref{table1}.
This similarity may be used as a starting heuristic for the 
phase-space-based preon-less interpretation of the rishon model,
involving in particular a further departure from 
the macroscopic dispersion relation (\ref{eq2}), 
which will be discussed shortly.
According to this interpretation, there is 
a correspondence between the dimensionality
of ordinary space and the number of colors. 
A somewhat different (relativity-inspired) version of this idea   
was proposed by Hidezumi Terazawa  who conjectured
that a {\it ``space-color correspondence
may become a clue to a possible relation
between fields (or matter) and space-time"}
\cite{Terazawa}.

\subsection{The Changing Meaning of ``Compositeness"}

The classical reductionist scheme is based on the concept of
a physical division of macroscopic 3D matter into
smaller and less massive 3D things.
This simple conception of compositeness cannot be used in 
the phase-space-based view of 
the HS model since the component rishons 
cannot be treated as 3D objects.
Indeed, the analogy existing between eg. the triplet of 
the generalized (``red" sector) momentum 
${\bf p}_R=(p_1, x_2, x_3)$) 
and the ordered set of three rishons $(TVV)$ 
suggests the association
of each of the three rishons with {\it one} dimension
of ordinary 3D space only.
With a single rishon 
(in a given ordered set of three) 
viewed as corresponding to
only one direction in the ordinary 3D world, 
the (3D) concept of spin  
cannot be assigned
to the rishon in question: thus, in the 
phase-space interpretation 
the individual rishons 
do not possess spin. 
By a similar analogy,
the set of three rishons (eg. $(TVV)$) cannot 
be rotated in 3D space and 
still remain ordered in the same way, ie. as $(TVV)$ (since 
some admixtures of $(VTV)$ and $(VVT)$ would have to appear - 
compare the situation for the triplets of generalized momentum 
in Table \ref{table2} ).
  
According to the SM, the lowest rung of the reductionist 
ladder involves leptons and confined quarks. 
 In line with our interpretation of the HS model,
going further ``down" (ie. below the lepton-quark rung) 
requires a bold departure 
from the remnants\footnote{The word "remnants" is used
 because quark confinement precludes the
applicability of Eq. (\ref{eq2}).}
of the macroscopic conception 
of ``real" matter 
- and points towards a completely different
understanding of the compositeness of matter-and-space.
Due to a change in the dimensionality 
assigned to rishons, an individual rishon 
cannot be thought  of as a ``thing",
and an additional rung of ``rishonian components" 
below that of leptons and quarks
cannot be thought of as composed of ordinary matter.
Thus, the process of a division of ``real" particles 
into smaller similar 3D stuff  
terminates at the lepton/quark level.
At the ``subsequent" level, 
the words ``to be composed of" totally
change their meaning \cite{Heisenberg}
and are understood as referring to the construction of
3D things from objects of lower dimensionality (1D).
Thus, in this approach {\it preons do not exist}.

For the benefit of a general reader
let me point out that
 in the phase-space-inspired version 
of the HS model the rishons 
 play the role quite analogous to that of regular polygons 
in the Plato's atomistic conception of 
matter\footnote{or to the scalene and 
isosceles right-angled triangles, 
considered by him to be the 
indivisible units from which
these polygons and further 
constructs may be built.}
\cite{Platoatoms}.
The emerging five regular polyhedrons (the Platonic solids) 
that in Plato's atomism correspond to 
the classical elements of fire, air, water, earth,  
and the Universe/aether
are then the analogues of leptons and quarks.

The above heuristic on the possible phase-space-related
meaning of the HS model and the one-dimensionality of rishons
may be replaced with more refined
mathematical arguments using Clifford algebra of
nonrelativistic phase space.
In particular, it can be shown that the structure of
the internal quantum numbers of a lepton-quark generation
naturally follows
from this algebra when the classical position and
momentum variables are replaced 
by the noncommuting quantum ones. 
For details (and the choice of even permutations), 
see \cite{Zen1},\cite{ZenCA}.

Our arguments in favour of the existence of a
close connection between matter and space have
led us to the language of phase space and an 
extension of the symmetry of reciprocity. 
In this language the appearance of 
the $U(1) \otimes SU(3)_C$ symmetry
and the pattern of lepton-quark generation 
are quite naturally explained by a link
to the macroscopic classical world. 
Technically speaking,  
the $U(1) \otimes SU(3)_C$ symmetry arises as a subgroup
of six-dimensional rotations 
in {\it classical} phase space.
With the physical principle underlying the
appearance of $U(1) \otimes SU(3)$ being
identified with the relevance of a phase-space-based
classical description of reality, I suspect
that some extension of this description 
should explain the appearance of $SU(2)_L$ as well.

  \section{Speculations}
According to the phase-space-inspired 
view of the rishon model, 
the lowest (lepton and quark) 
rung of the reductionist ladder involves
a loss of macroscopic spatial separability 
and rotational symmetry. 
 Indeed, 
in the permutation-induced 
classical description, the individual 
colored ``objects" are associated with 
rotationally and translationally
non-invariant analogues of expression (\ref{eq2}), 
in which ${\bf p}^2$ is replaced by 
expressions such as
\begin{equation}
\label{2nonrota}
{\bf p}^2_R=p_1^2+x_2^2+x_3^2. 
\end{equation}
Eq.(\ref{2nonrota}) leads 
 to a further departure from (\ref{eq2}), ie.
to a rotationally and translationally
non-invariant generalisation of 
the classical dispersion relation. 
For the red sector it reads
\begin{equation}
\label{REDdisprel}
E^2= {\bf p}^2_R+m^2.
\end{equation}

One may discard such formulas 
together with the phase-space interpretation 
of the HS model as unapplicable to the real 
rotationally covariant macroscopic world.                            
Yet, a different, speculative point of view is also possible. 
According to this alternative view (to which I subscribe),
the {\it individual} ``colored objects" may exhibit 
rotationally non-invariant features
and still be considered physically acceptable   
if only they can conspire to form rotationally covariant
 composite systems at the more macroscopic 
level. 

Now, restoration of rotational covariance 
requires the cooperation of three sectors of objects
satisfying three differently modified dispersion relations. 
These are the sectors of ``red", ``green", and ``blue" 
objects with their generalized momenta being
${\bf p}_R=(p_1, x_2, x_3)$, ${\bf p}_G=(x_1, p_2, x_3)$, 
${\bf p}_B=(x_1, x_2, p_3)$, that together 
make the formation of 
standard vectors (like $(p_1,p_2,p_3)$) possible.
Similarly, restoration of rotational covariance 
at the rishon level should require
the cooperation of the corresponding 
rishon triplets $(TVV)$, $(VTV)$, $(VVT)$,
ie. the cooperation of three differently colored quarks.
Furthermore, due to the position dependence of the
generalized dispersion relations (eg. Eq.(\ref{REDdisprel})),
the individual colored quarks do not possess the classical 
reductionist feature of spatial separability,
and require cooperation of other quarks (or antiquarks)
in the formation of translationally invariant 
(string-like/flux-tube) structures.

In brief, it is the non-standard (``confining") 
nature of dispersion relations
like (\ref{REDdisprel})
 that is hidden behind the non-existence
of individual colored objects in our macroscopic, classical world.
Furthermore, it is the requirement of a proper rotational
and translational behaviour, 
imposed upon the underlying quark structures,
that is supposed to lie at the origin of the emergence 
of ordinary 3D space.
In our view, the rudiments of 
the emerging matter-defined 3D space
are formed via the construction 
of colorless hadron-level structures.

It is at the hadronic level, 
when  - with the help of quark conspiracy - 
the effective 3D point
may be successfully constructed, 
that a talk about the extension to
special relativity could be started.
Yet, the absence of $c$ in the successful 
formulas for both hadronic scales (mass and string slope, Eqs.(\ref{mH},\ref{alphaprim}))
seems to indicate the relative unimportance of 
special relativity at the deep quantum 
level.\footnote{\label{SRbackground} 
It should be stressed that
 the continuous relativistic spacetime, used 
as a background in the SM,  constitutes 
a theory-based idealized extrapolation of 
a concept defined by Einstein with the help of
{\it macroscopic} objects (clocks and rods)  
\cite{Zimmerman}.
It  is not  matter-defined at 
the microscopic quantum level since 
atomic-, hadronic-, or quark-size rods and clocks 
do not exist \cite{Wigner}. 
Yet, as argued above, 
the world of quarks  
 does possess properties that seem to allow 
the construction (emergence) of the rudiments 
of  hadron-defined 3D space.}

We viewed the non-standard form of the  
modified dispersion relations
(\ref{REDdisprel})  as expressing  
in the
phase-space-induced language
the stringy nature of quark
confinement. 
This form suggests
that QCD - with its input of 
conceptually standard dispersion
relations for quarks and universal continuous
background spacetime - 
 provides 
a significantly simplified and idealized description of the long-distance
 confining fluxtube aspects of inter-quark forces.
Although all our theories are  
idealizations or approximations,
the question emerges how 
QCD could generate rotationally 
noninvariant fluxtubes that
would correspond to the stringy 
formulas (\ref{2nonrota},\ref{REDdisprel}) of our scheme. 
I believe we could learn this from studies of long-distance
properties of hadrons.
 In fact, I suspect that in its treatment of 
the confined behaviour of quarks
  QCD misses something that does not appear at short distances.
There are phenomenological hints from baryon spectroscopy
that it may be so \cite{CapstickRoberts}.
Namely, the standard constituent quark model
and lattice QCD both predict the existence of 
many more (by a factor of 2 or so) 
excited baryonic states than seen
experimentally. 
The continuing absence of these states 
from the observed spectrum 
(see \cite{Chen2019}) 
may indicate
that in excited baryons some internal spatial degrees 
of freedom are missing, thus hinting again
at the connection of strong interactions 
with the emergence (or construction) of space \cite{Zen2022}.

\section{Synopsis}

1) The analogy  between the phase-space-permutation-induced
pattern of the sets of generalized momenta
and the rishon structure of leptons and quarks
leads to the interpretation of rishons
as one-dimensional ``objects". Consequently, rishons
should not be viewed as subparticles, and the HS model
should not be regarded as a preon model.
 
\noindent
2) Our view of the HS model associates the structure 
of a single generation of leptons and quarks with
the appearance of $2 \times (1+3)$ ways in which sets of
generalized momenta (or positions) appear in a matter-space
symmetric approach.

\noindent
3) This view
 provides also a classical macroscopic reason
for the appearance of the $U(1) \times SU(3)_C$ symmetry 
and suggests a simple interpretation for the origin
of the non-existence of individual colored quarks.
It supports the idea that the rudiments of ordinary 3D
space begin to emerge through 
the formation of hadrons.

\vfill

\vfill

\end{document}